# Reconfigurable sub-μm spin-wave majority gate with electrical transducers

Giacomo Talmelli[1,2], Thibaut Devolder[3], Nick Träger[4], Johannes Förster[4], Sebastian Wintz[4,5], Markus Weigand[4,6], Hermann Stoll[4,7], Marc Heyns[1,2] Gisela Schütz[4], Iuliana P. Radu[1], Joachim Gräfe[4], Florin Ciubotaru[1], and Christoph Adelmann[1,*]

## Affiliations

[1]Imec, 3001 Leuven, Belgium.

[2]KU Leuven, Departement Materiaalkunde, SIEM, 3001 Leuven, Belgium.

[3]Centre de Nanosciences et de Nanotechnologies, CNRS, Université Paris-Sud and Université Paris-Saclay, 91120 Palaiseau, France.

[4]Max-Planck-Institut für Intelligente Systeme, 70569 Stuttgart, Germany.

[5]Paul Scherrer Institut, 5232 Villigen, Switzerland.

[6]Helmholtz-Zentrum Berlin, 12489 Berlin, Germany

[7]Institut für Physik, Johannes Gutenberg-Universität Mainz, 55099 Mainz, Germany

*Correspondence to: christoph.adelmann@imec.be

## Abstract

Spin waves are excitations in ferromagnetic media that have been proposed as information carriers in hybrid spintronic devices with much lower operation power than conventional charge-based electronics. Their wave nature can be exploited in majority gates by using interference for computation. However, a scalable spin-wave majority gate that can be co-integrated alongside conventional electronics is still lacking. Here, we demonstrate a sub-μm inline spin-wave majority gate with fan-out. Time-resolved imaging of the magnetization dynamics by scanning transmission x-ray microscopy illustrates the device operation. All-electrical spin-wave spectroscopy further demonstrates majority gates with sub-μm dimensions, reconfigurable input and output ports, and frequency-division multiplexing. Challenges for hybrid spintronic computing systems based on spin-wave majority gates are discussed.

## INTRODUCTION

Spin waves are dynamic excitations in ferromagnetic media with characteristic wavelengths from nm to μm scales and frequencies from GHz to THz. Due to their low intrinsic energy, they have received increasing interest as information carriers in spintronic computation and communication schemes (1–8) operating potentially at much lower power than current charge-based complementary metal-oxide-semiconductor (CMOS) technology (9). Recently, the concept of a hybrid spin-wave–CMOS computer has been proposed that combines spin-wave-based logic circuits with a CMOS periphery (8,9). The logic circuits themselves take advantage of the wave nature of spin waves and use interference for computation (7,8). The communication between individual spin-wave circuits and the outside world, including memory, is obtained by conventional CMOS electronics. Transducers provide signal conversion at the interfaces between charge- and spin-wave-based subsystems.

Spin waves are particularly suited for the realization of compact interference-based majority gates (7,10,11). Majority gates have recently regained intense interest since they promise circuit simplification with respect to conventional designs with potential for area reduction even with relaxed critical device dimensions with respect to CMOS transistors. The relaxed dimensions result in simplified lithography requirements and can be expected to reduce cost. In initial work, spin-wave majority gates (SWMGs) based on a trident design have been proposed (7,11) and their basic functionality has been experimentally demonstrated for 100-μm- to mm-size devices (12–14). To become competitive with current CMOS technology in terms of circuit area, SWMGs need to reach critical dimensions of the order of 100 nm (7). However, the trident design has severe drawbacks when miniaturized to the nanoscale, such as narrow operation windows, strong spin-wave attenuation at the bends, and limited compatibility with conventional lithography processes. Moreover, the previous mm-scale experimental demonstration has relied on single-crystal $Y_3Fe_5O_{12}$ (YIG), resulting in low spin-wave group velocities below 1 μm/ns in nm-thin films, long spin-wave lifetimes, and low computational throughput (see Table 1). By contrast, an ideal SWMG combines compact size, scalability to target dimensions, fast operation, possibility for fan-out, and a wide operation window with a flexible design that alleviates lithography constraints. In this paper, we describe a scalable sub-μm SWMG based on the ferromagnetic metals permalloy ($Ni_{80}Fe_{20}$) and $Co_{40}Fe_{40}B_{20}$ that can be fabricated using CMOS-compatible processes. Inductive antennas are used as spin-wave transducers at the input and output ports of the devices. The results indicate that spin waves can be used as information carriers and for computation in scaled logic gates. Remaining challenges in circuit and transducer technologies are also discussed below.

# RESULTS

## Operation principle of the inline spin-wave majority gate

The basic structure and the operation principle of our inline SWMG are depicted in Fig. 1A. Three input ports, $P_1$, $P_2$, and $P_3$, as well as one or two output ports, O and O′, are placed at equally-spaced positions on a spin-wave waveguide (15), *i.e.* at positions $n \times 2F$ with $n = 1, 2, 3, 4$ and $F$ the characteristic dimension of the device. For transducers with an area of $F^2$ at the ports, the inline SWMG occupies an area of 16 $F^2$ (20 $F^2$ for a device with two output ports and thus a fan-out of two), much smaller than CMOS implementations and even significantly more compact than trident-based SWMG designs. Using microwave signals at the input ports, coherent spin waves of unit amplitude are launched at the three input ports and propagate in the waveguide. Binary logic signals are encoded in the phases of the individual spin waves using phases of 0 and π as logic 0 and 1, respectively. Constructive or destructive interference leads then to an output wave with a phase that corresponds to the majority of the individual spin-wave phases (Fig. 1B). The amplitude of the output wave contains further information whether weak or strong majority is obtained. The generalization to majority gates with more than three inputs is straightforward. The operation of the SWMG requires that the phases of the individual spin waves for a given logic level are matched at the output port. In spin-wave circuits, it is desirable to use identical microwave signals for a given logic level at all input ports, which is realized under "resonant" operation conditions when the interport distance $2F$ is equal to $N \times \lambda$ with $N = 1, 2, 3,\ldots$ and $\lambda$ the spin-wave wavelength. The spin-wave phase then rotates by integer multiples of 2π during propagation from any input to the output port, leading to matched phases both at the individual inputs and the output.

## Implementation

Miniaturizing SWMGs to nm dimensions allows for the usage of ferromagnetic waveguide materials with moderate Gilbert damping (and thus shorter spin-wave attenuation lengths) than ultralow-damping single-crystal YIG that has been used for mm-size trident-based SWMG realizations (12). YIG suffers from temperature-dependent low saturation magnetization and low spin-wave group velocities (16) (see Table 1). Moreover, single-crystal YIG cannot be integrated epitaxially in hybrid systems alongside conventional CMOS on Si wafers due to the large lattice mismatch. By contrast, metallic ferromagnets such as CoFeB and permalloy ($Ni_{80}Fe_{20}$) with much larger saturation magnetization and higher Curie temperature promise faster operation, lower temperature sensitivity, and are compatible with established semiconductor process technology. Our implementation will thus be based on $Co_{40}Fe_{40}B_{20}$ and permalloy waveguides with widths down to 850 nm (Fig. 1C). In such narrow waveguides, the mode patterns of confined spin waves deviate significantly from the plane waves employed in previous macroscopic SWMG implementations (12) due to the nonuniformity of the static magnetization and the internal effective field in the commonly-used Damon-Eshbach configuration with the magnetization oriented transverse to the spin-wave waveguide (17). This configuration leads to spin waves with large group velocities (of the order of 5–10 μm/ns for our $Co_{40}Fe_{40}B_{20}$ waveguides). Micromagnetic simulations in Fig. 1D for 850 nm wide $Co_{40}Fe_{40}B_{20}$ waveguides in the Damon-Eshbach geometry show the excitation of spin waves confined in the center of the waveguide and modulated by backward-volume spin waves that are excited at the edges and propagate preferentially towards the center. Yet, the simulations demonstrate majority-gate operation in such a device despite the rather complex mode patterns when the phase of the magnetization precession at the output port is analyzed. Animations of the full magnetization dynamics for different sets of input phases can be found in the supplementary information.

In hybrid spin-wave–CMOS systems, transducers play an important role as they transform signals between spin-wave and charge domains. Different approaches for transducers that couple spin-wave and microwave signals have been reported in the literature (18,19). In this work, we employ inductive microwave antennas (20) at the input and output ports because they combine high maturity and robustness with broadband excitation and detection of spin waves (21) in the Damon-Eshbach geometry. For all-electrical operation of SWMGs, U-shaped antennas were used as spin-wave transducers because of their low parasitic crosstalk. By contrast, SWMG imaging by time-resolved scanning transmission x-ray microscopy was performed using more compact single wire antennas. Details about the antenna design can be found in the supplementary information.

**Time-resolved imaging of spin-wave majority gate operation**.

The operation of an inline SWMG can be visualized by imaging the magnetization dynamics in the waveguide by time-resolved scanning transmission x-ray microscopy. In these experiments, spin waves are excited in a 2.0 μm wide permalloy waveguide by microwave currents in three input antennas $P_1$, $P_2$, and $P_3$, each separated by distances of $2F$ = 2.5 μm. Binary logical signals 0 and 1 are encoded as spin-wave phases of 0 (reference) and π, respectively. The logical output signal is determined by extracting the phase of the resulting spin wave at position O, about 2.2 μm away from $P_1$, from the measured time dependence of the magnetization dynamics. An operation frequency of 8.6 GHz and an external magnetic bias field of $\mu_0 H_{\text{ext}}$ = 80 mT lead to a measured spin-wave wavelength of 2.4 μm, which is very close to the interport distance. Thus, resonant operation conditions are approximately realized with $N$ = 1. The phase maps of the magnetization dynamics at position O in Fig. 2 lead to a set of output phases corresponding to the majority gate truth table for all combinations of logical input signals. Animations of the magnetization dynamics for selected combinations of input phases can be found in the supplementary information.

Figure 2 and the animations in the supplementary information show that spin waves do not only propagate towards the chosen output port position but also along the waveguide in the opposite direction. Analyzing the spin-wave phase at output O′ can also be used to extract the majority of the input phases. The inline SWMG in resonant operation conditions thus allows for a fan-out of two with only a small additional area of 4 $F^2$ (total area 20 $F^2$) and without the need to convert the spin-wave signal back into the microwave domain. This is a highly desirable property for the design of more complex spin-wave circuits. Adding additional output ports at other locations where the output waves are in phase can increase the fan-out even more. Inverting output ports are also possible at positions where the spin waves accumulate an additional phase shift of π. We note that the observed nonreciprocity of the spin-wave intensity is a consequence of the chirality of the exciting magnetic field generated by the inductive antennas and can be avoided by using other types of spin waves, such as forward-volume spin waves, or nonchiral spin-wave transducers, *e.g.* magnetoelectric transducers (22,23). However, as long as spin-wave attenuation is small, the nonreciprocity does not affect device operation and fan-out.

**All-electrical spin-wave engineering.**

Building spintronic logic gates based on interference requires the quantitative assessment of the spin-wave properties such as their dispersion relation and their propagation loss to allow for the control of both the amplitude and the phase of the spin waves at the output port. This can be achieved by a series of all-electrical two-port microwave measurements, in which spin waves are excited at an input port and, after propagation, analyzed with phase sensitivity at the output port. The phase-sensitive spin-wave transmission can then be inferred from the bias-field derivative of the microwave $S_{21}$-parameter, $dS_{21}/dH_{ext}$, as shown in Fig. 3A for a 850 nm wide $Co_{40}Fe_{40}B_{20}$ waveguide using $P_3$ as input and O as output port at a distance of 6.9 μm. The shape of the transmitted signal can be understood as follows: during propagation over a distance $r$ between input and output port, the spin-wave phase rotates by a factor $e^{ikr}$, with $k = 2\pi/\lambda$ the spin-wave wavevector. As a result, the real and imaginary parts of $dS_{21}=dH_{ext}$ oscillate both in $r$ and $k$. The correspondence between the wavevector $k$ of the spin wave and its frequency $f$ above the ferromagnetic resonance frequency where $k$ vanishes is given by the spin-wave dispersion relation. Spin waves attenuate during propagation due to their finite spin-wave lifetime (typically 1 to 1.5 ns for permalloy and $Co_{40}Fe_{40}B_{20}$ waveguides in the experimental frequency range), leading to a decay of the spin-wave intensity with propagation distance. Typical spin-wave group velocities in such waveguides in the studied frequency range are 5-10 μm/ns, resulting in decay lengths of the order of 10 μm, larger than characteristic propagation distances in the studied devices. The envelope of the frequency dependence of the spin-wave response is determined mostly by the $k$-dependent coupling efficiency of the antennas. A one-dimensional model that takes into account the dispersion relation (24), the attenuation, and the antenna coupling efficiency is in excellent agreement with the experiment (see the supplementary information). The extracted spin-wave dispersion relations for 850 nm wide and 30 nm thick $Co_{40}Fe_{40}B_{20}$ waveguides at different magnetic bias fields are shown in Fig. 3B. At weak magnetic fields below the saturation of the magnetization in the transverse direction (< 50 mT), contributions from backward-volume spin-wave modes are also visible in the experimental data in Fig. 3A. However, these modes possess very short decay lengths and can therefore not be used for SWMG operation.

The two-port all-electrical microwave measurements also provide information about the signal levels that can be obtained with our setup. Since spin waves are low-energy excitation of the ferromagnets, the power transmission is rather weak (see supplementary information). For logic applications and interfacing with CMOS circuits, voltage signal levels are however more appropriate. Figure 3C shows the voltage signal transmission ratio ($V_{in}/V_{out}$) deduced from measured antenna impedances and micromagnetic simulations. For waveguides with widths in the (sub)-μm range and our antenna design, transmission ratios are of the order of a few 0.1 to 1%. Further improvements can be achieved by reduced spin-wave attenuation in scaled devices and lower antenna resistances. Furthermore, an optimization of the matching conditions in microwave periphery may also increase signal transmission levels. However, since this would also reduce the available bandwidth, it was not implemented in our wide-band experiments. Ultimately, magnetoelectric transducers promise to increase the voltage transmission ratios even further (7,22).

### Electrical operation of sub-μm spin-wave majority gates.

To design scalable SWMGs, three different dimensions need to be considered: the waveguide width, the interport spacing, and the spin-wave wavelength. We first consider an 850 nm wide $Co_{40}Fe_{40}B_{20}$ waveguide with U-shaped antennas and an interport spacing of $2F$ = 2.3 μm. Selecting a spin-wave wavelength of $\lambda$ = 1.15 μm, *i.e.* resonant conditions with $N$ = 2, leads to higher spin-wave excitation and detection efficiency by the U-shape antennas in this wavelength range with respect to $N$ = 1 and is therefore favorable. At the corresponding frequency of 13.86 GHz (at a transverse magnetic bias field of $\mu_0 H_{ext}$ = 90 mT, see Fig. 3B), we explore all combinations of logical levels (input phases of 0 or π) and extract the phase at the output by analyzing both the real and imaginary parts of the $S_{21}$-parameter (Fig. 4A) to successfully construct the full logic truth table of the majority function. All output states in this SWMG can be clearly distinguished within a frequency band of about 250 MHz around the target frequency (Fig. 4B). The clear separation between levels including weak and strong majority cases indicates that the device concept can be extended to $n$-input SWMGs with $n$ > 3 by adding additional input ports. We note that the large bandwidth is a consequence of the high saturation magnetization of $Co_{40}Fe_{40}B_{20}$ and the steep spin-wave dispersion relation at the chosen conditions. By contrast, using YIG as waveguide material leads to a much narrower operational frequency band of only about 15 MHz (see Table 1). Due to the low Curie temperature of YIG and the resulting strong dependence of the magnetization on temperature, this results in a narrow temperature range, in which SWMG operation is possible at a given frequency. By contrast, operation using $Co_{40}Fe_{40}B_{20}$ waveguides is possible up to much higher temperatures without the need to adjust the operation frequency (see Table 1).

Due to its symmetry, the inline SWMG can be used in a flexible way and any port can be chosen to be the output. Figures 4C and 4D illustrate two possible configurations: a "conventional" configuration with one of the outer antennas as the output port (Fig. 4C), or an alternative second implementation when the output is one of the inner antennas (Fig. 4D). Using a 4.7-μm-wide $Co_{40}Fe_{40}B_{20}$ waveguide with an interport spacing of $2F$ = 2.3 μm, the experimental spin-wave transmission signals for these configurations demonstrate that the SWMG works equally well within a frequency band of over 300 MHz when the roles of the ports are exchanged. The behavior is affected in this device design by the nonreciprocity of spin-wave excitation in the Damon-Eshbach geometry and thus the spin-wave intensities need to be adjusted when changing configurations. However, as mentioned before, this issue can be avoided by using other types of spin waves, such as forward-volume spin waves, or nonchiral spin-wave transducers.

**Frequency-division multiplexing and operation at sub-μm wavelengths.**

The usage of waves for computation allows for frequency-division multiplexing and parallel computation in a majority gate with fixed geometry, enabling a larger computational throughout without additional area consumption. At low excitation powers, spin waves are noninteracting and different frequency channels can thus be used independently for logic operations. When nonresonant conditions are employed, SWMGs can work at any frequency in the spin-wave band above the ferromagnetic resonance, and the spacing of individual frequency subbands is only limited by the intrinsic line broadening due to the finite spin-wave lifetime (about 100 to 150 MHz for the ferromagnetic materials chosen in this study). When resonant operation conditions with matched interport distances and spin-wave wavelengths are chosen, frequency-division multiplexing can use a series of harmonics with $\lambda = 2F/N$ and $N$ = 1, 2, 3, .... The usage of different harmonics in a SWMG is illustrated by the calculated frequency dependence of the output signal (Fig. 5A) for all input phase combinations and the dispersion relation in Fig. 5B for a 4.7 μm wide $Co_{40}Fe_{40}B_{20}$ waveguide and a magnetic bias field of $\mu_0 H_{ext}$ = 42 mT. For an interport spacing of 2.3 μm, resonant operation of the SWMG with $N$ = 1 and $N$ = 2 is realized at frequencies around 12.2 GHz and 15.1 GHz, respectively, as indicated in Fig. 5A. This behavior is experimentally confirmed at frequencies of 12.22 GHz and 14.92 GHz, as shown in Figs. 5C and 5D, respectively. The experimental logic signals indicate that the majority function is obtained in the same device at these frequencies with $N$ = 1 and $N$ = 2, respectively.

Additional higher harmonics with shorter wavelengths are outside the frequency range of the experimental set-up. However, lowering the magnetic bias field to $\mu_0 H_{ext}$ = 12 mT brings spin waves with sub-μm wavelengths into the accessible frequency range (Fig. 5B). SWMG operation at a wavelength as low as 600 nm in a device using a 4.7 μm wide $Co_{40}Fe_{40}B_{20}$ waveguide and an interport spacing of 1.2 μm is demonstrated in Fig. 5E at a frequency of 17.72 GHz for $N$ = 2. The experimental logic signals allow for the deduction of the full majority function including a clear distinction between strong and weak majority. Operation at smaller wavelengths is possible using a set-up with larger frequency range as well as smaller antennas. Micromagnetic simulations do not indicate any fundamental limit for SWMG operation at wavelengths of the order of 100 nm and beyond (7).

### Speed of operation.

The operation speed of an inline SWMG is determined by several factors. The propagation time of spin waves from the inputs to the output is one of the contributions to the time required for a computation. For $Co_{40}Fe_{40}B_{20}$, the group velocity of spin waves with wavelengths of the order 1 μm is about 5 μm/ns (see Table 1). Hence, the propagation in our devices contributes about 1 ns to the computation time. It is clear that further device scaling can reduce the contribution into the sub-ns range.

Different computations with different logic input values require also the change of the phase of the input microwave signals. Spin waves do not instantaneously follow the change of an input signal phase but require some time for phase resynchronization. The resynchronization time can be assessed by micromagnetic simulations (Fig. 6A). In these simulations, the phase of the excitation field generated by the antenna at input port $P_2$ is abruptly changed at 1.0 ns from 0 to π, whereas the phase of the wave propagating from $P_1$ remains π. Monitoring the phase of the spin wave at the output position O after interference (Fig. 6B) indicates that the new computation result is obtained after about 700 ps, which includes both the propagation delay the and resynchronization time. These results indicate that spin waves resynchronize with external signals in less than 500 ps and demonstrate the potential of scaled SWMGs to operate at phase modulation frequencies of 2 GHz and above.

## DISCUSSION

These results indicate that SWMGs can be key elements of a future hybrid spin-wave–CMOS computing platform (7–9). In this concept, computation is performed in spin-wave circuits with low power consumption whereas CMOS-based devices provide logic input signals, long distance data exchange, as well as data storage. Recent benchmarking of the performance of such systems has suggested that they bear promise for ultralow power computation with comparable areas as conventional CMOS (9,25). While such circuits may not reach computation delays and the operation speed of high-performance CMOS despite the fast operation demonstrated by micromagnetic simulations and reported above, it has nonetheless been shown that hybrid spin-wave–CMOS can outperform conventional CMOS at the 10 nm node in the area-delay-power product (9). Frequency-division multiplexing as demonstrated above may allow for a trade-off between computational throughput and power or area, which are increased by the additional necessary periphery.

To obtain a functionally complete set of logic spin-wave gates to design the spin-wave circuits at the heart of hybrid spin-wave–CMOS systems, SWMGs must be complemented by another wave-based logic gate. Inverters (logic NOT) provide a highly appealing solution since their implementation in phase-coded wave-based computing corresponds to a simple phase shift by π. In contrast to CMOS technology, wave-based inverters do not need to be separate logic gates but can be integrated in the majority gate design. Physical implementations can be based on a spin-wave delay line with a length of $(i - 1/2) \times \lambda$ , with $i$ = 1, 2, 3, ... an integer and $\lambda$ the spin-wave wavelength, which can be *e.g.* realized by shifting input or output ports away from their equally-spaced positions. In addition, shifting the phase or reverting the polarity (signal *vs.* ground) of an input or output port can also be used to invert the logic signal. These implementations lead to little or no increase in device area, and, in the second case, to no additional restrictions on operating conditions. The time-resolved scanning transmission x-ray microscopy imaging of SWMG operation visualizes this possibility to integrate inverters in the device since it shows that the phase of the output wave oscillates in space. The demonstration of the calculation of the minority (inverted majority) and the majority function in the same device by adjusting input phases is also demonstrated in the supplementary material (Fig. S6).

By combining majority gates and inverters, every logic circuit can be synthesized. The demonstration of electrically driven scalable SWMGs and inverters at the sub-μm scale with a large operational window that can be manufactured using CMOS-compatible techniques establishes a proof-of-concept of spin-wave computing. Reaching device areas that are competitive with CMOS requires the scaling of the devices to dimensions of the order of 100 nm (9). Despite the experimental challenges, no fundamental obstacle exists to reach this target. It should be noted that the critical dimensions of spin-wave devices at equal area are almost one order of magnitude larger than current CMOS implementations. This enables manufacturing using much simpler lithography and patterning processes and promises much reduced cost. Beyond individual logic gates, implementations of spin-wave circuits will require additional devices, such as spin-wave-switched nanomagnets (7,26), spin-wave amplifiers (27), or magnonic directional couplers (28,29) that can be used for signal restoration within the spin-wave circuit (8,29). Recently, a proof-of-concept of magnonic directional couplers has been realized experimentally using YIG (30).

Achieving much lower computation power than current CMOS technology will also require the usage of spin-wave transducers with higher signal transfer efficiency. While the used inductive antennas are highly mature and robust and allow for broadband excitation and detection of spin waves, alternative concepts may be required, such as magnetoelectric transducers (7–9). Magnetoelectric transducers operate directly using voltage signals and do not share the limitations of inductive antennas. First results of the generation and detection of spin-waves by magnetoelectric transducers have already been reported (22) although the scalability of the approach has not been demonstrated yet.

## Materials and Methods

***Materials and devices.*** The devices consisted of ferromagnetic waveguides with widths between 850 nm and 4.7 µm. Waveguide lengths were typically about 30 µm, sufficiently long to avoid the influence of spin-wave reflection at the waveguide ends on the device behavior. Ta/$Ni_{80}Fe_{20}$/Ta, ($Ni_{80}Fe_{20}$ = permalloy, 3 nm/30 nm/3 nm) waveguides were used for the scanning transmission x-ray microscopy measurements whereas Ta/$Co_{40}Fe_{40}B_{20}$/Ta (3 nm/30 nm/3 nm) was used in all-electrical microwave experiments. Spin waves were excited and detected by inductive antennas made from Al for scanning transmission x-ray microscopy and Au for microwave measurements, electrically connected to coplanar microwave waveguides. The saturation magnetization of the films was determined by vibrating-sample magnetometry and was 0.8 MA/m and 1.36 MA/m for $Ni_{80}Fe_{20}$ and $Co_{40}Fe_{40}B_{20}$, respectively. Gilbert damping parameters were determined by ferromagnetic resonance measurements to be $\alpha = 7\times10^{-3}$ for Ta/$Ni_{80}Fe_{20}$/Ta and $\alpha = 4\times10^{-3}$ for Ta/$Co_{40}Fe_{40}B_{20}$/Ta.

The devices for time-resolved scanning transmission x-ray microscopy have been fabricated on commercially available 100 nm thick square $SiN_x$ membranes (area of 100×100 µm$^2$) fabricated on $SiN_x$/Si (100). A combination of e-beam lithography and lift-off processing was used to pattern a sputtered Ta/$Ni_{80}Fe_{20}$/Ta (3 nm/30 nm/3 nm) ferromagnetic waveguide on top of the membrane. Subsequently, a 60 nm $SiO_2$ insulating layer was sputtered over the wafer, then e-beam lithography and lift-off were used to pattern Ti/Al (10 nm/100 nm) inductive microwave single-wire antennas and coplanar waveguides for electrical contact. Al metallization was chosen to avoid strong x-ray absorption during the measurements so that spin waves could be detected also underneath the antennas.

Devices for all-electrical characterization were processed on $SiO_2$/Si substrates. A Ta/$Co_{40}Fe_{40}B_{20}$/Ta (3 nm/30 nm/3 nm) stack was sputter deposited and subsequently annealed at 300°C for 10 minutes in an $N_2$ environment. Next, a sputtered 60 nm thick $SiO_2$ layer was deposited as an electrical insulation for the antennas and as a hard mask for the patterning process. E-beam lithography and inductively coupled plasma etching were then used to define the waveguide in the hard mask layer, which was finally transferred into the ferromagnetic Ta/$Co_{40}Fe_{40}B_{20}$/Ta stack by ion beam etching. Subsequently, an about 300 nm thick planarization layer (spin-on-carbon) was spun on the structure and recessed by inductively coupled plasma etching to reduce the step height after etching, followed by a 30 nm thick sputtered $SiO_2$ layer to improve the electrical insulation. E-beam lithography and lift-off were finally used to fabricate Ti/Au (10 nm/100 nm) coplanar waveguides (for electrical contact) and U-shaped microwave antennas. The gap of the U-shaped antennas was defined by additional e-beam lithography and ion beam etching steps to reduce the risk of shorting.

***Time-resolved scanning transmission x-ray microscopy.*** Time-resolved scanning transmission x-ray microscopy measurements were carried out at the MAXYMUS end station at the UE46-PGM2 beamline at the Bessy II synchrotron within the Helmholtz-Zentrum Berlin. The samples were illuminated under perpendicular incidence by circularly polarized light in an external in-plane magnetic bias field of up to $\mu_0 H_{\text{ext}}$ = 240 mT that was generated by a set of four rotatable permanent magnets (31). The photon energy was set to the absorption maximum of the Fe $L_3$ edge to get optimal contrast for imaging. A lock-in-like detection scheme allowed for the measurement of the magnetization dynamics—and in particular spin waves—excited at microwave frequencies with a time resolution of 50 ps using all photons emitted by the synchrotron. Input signals were provided by an arbitrary waveform generator, allowing for the independent control of both amplitude and phase of multiple microwave excitation channels. Further details on the detection and data analysis scheme can be found in (32).

***All-electrical microwave measurements.*** All-electrical measurements of SWMGs were performed using a two-port Vector Network Analyzer (VNA). The signal from port 1 of the VNA was split into three separated lines connected to three ground-signal-ground microwave probes connected to the three input coplanar waveguides attached to the input antennas, $P_1$, $P_2$, and $P_3$, of the SWMG. Two power splitters were used as schematically depicted in Fig. S1. In all lines, variable attenuators and phase shifters were inserted such that it was possible to control the amplitude and phase of each input independently, especially to match amplitudes of the spin waves at port O and create the two logic phase states, 0 and π. The output O was connected to port 2 of the VNA, allowing for a phase-sensitive measurement of the transmitted power. The bandwidth (frequency cutoff) of the entire circuit was 18 GHz. During all-electrical measurements, the power delivered to individual antennas was -6 dBm (~0.25 mW). Considering the measured antenna impedances, this corresponded to an input voltage of about 120 mV.

***Micromagnetic simulations.*** Micromagnetic simulations have been performed using the object-oriented micromagnetic framework (OOMMF) software package (33). The geometry and the magnetic material parameters were chosen to correspond to experimental SWMGs with $Co_{40}Fe_{40}B_{20}$ waveguides. The saturation magnetization was assumed to be 1.36 MA/m, the exchange stiffness constant was 18.6 pJ/m, the Landé $g$-factor was 2.07 (34), and the Gilbert damping was $\alpha = 4\times10^{-3}$. These parameters led to dispersion relations in excellent agreement with the experiment as shown in the supplementary text below.

## Acknowledgments

**General**: The authors would like to thank the Helmholtz-Zentrum Berlin (HZB) for the allocation of synchrotron radiation beamtime and Sina Mayr (PSI) for her support of the time-resolved scanning transmission x-ray microscopy measurements. Odysseas Zografos (imec) and Sorin Cotofana (TU Delft) are acknowledged for many valuable discussions on spin-wave circuits. **Funding:** This work has been supported by imec's industrial affiliate program on beyond-CMOS logic. It has also received funding from the European Union's Horizon 2020 research and innovation program within the FET-OPEN project CHIRON under grant agreement No. 801055 as well as under grant agreement No. 730872. **Author contributions:** G.T, T.D, M.H., G.S., I.P.R., J.G., F.C., and C.A. conceived the experiments and designed the samples. G.T. fabricated the samples with device integration support from I.P.R and F.C. T.D. developed the experimental microwave set-up. G.T. and T.D. performed the microwave measurements. N.T., J.F., S.W., M.W., H.S., G.T., and J.G. performed the time-resolved scanning transmission x-ray microscopy measurements. G.T., T.D., F.C., and C.A. developed the dispersion models, confirmed by F.C. by micromagnetic simulations. G.T., T.D., N.T., S.W., M.H., G.S., J.G., F.C., and C.A. analyzed and interpreted the data. G.T., T.D., J.G., F.C., and C.A. prepared the manuscript. All authors commented on the manuscript. **Competing interests:** The basic device concept of the inline SWMG gate has been patented in European patent No. EP3339871. **Data and materials availability:** All data supporting the findings of this study are available within the article and its supplementary information. Extra data are available from the corresponding author upon request.

## Figures and Tables

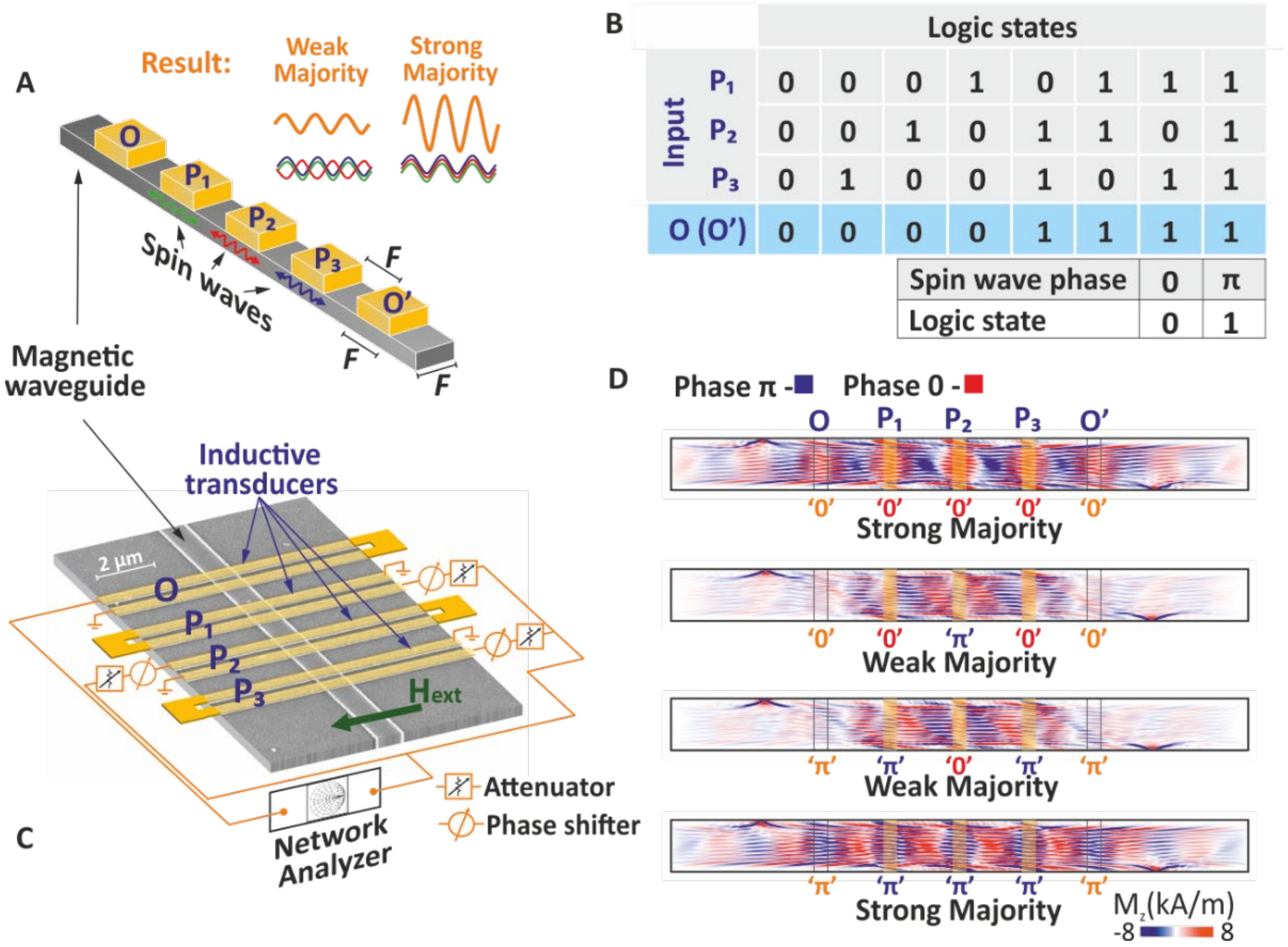

| | Logic states | | | | | | | |
|---|---|---|---|---|---|---|---|---|
| Input $P_1$ | 0 | 0 | 0 | 1 | 0 | 1 | 1 | 1 |
| Input $P_2$ | 0 | 0 | 1 | 0 | 1 | 1 | 0 | 1 |
| Input $P_3$ | 0 | 1 | 0 | 0 | 1 | 0 | 1 | 1 |
| O (O′) | 0 | 0 | 0 | 0 | 1 | 1 | 1 | 1 |

| Spin wave phase | 0 | π |
|---|---|---|
| Logic state | 0 | 1 |

**Figure 1. Device structure and operation principle of the inline spin-wave majority gate. A.** Schematic of an inline SWMG with three inputs ($P_1$ to $P_3$) and two outputs (O and O′), *i.e.* with a fan-out of two. **B.** Truth table of the majority function. **C.** Scanning electron micrograph of a SWMG with an 850 nm wide $Co_{40}Fe_{40}B_{20}$ waveguide. **D.** Steady-state snapshots of the out-of-plane magnetization obtained by micromagnetic simulations of a SWMG using an 850 nm wide $Co_{40}Fe_{40}B_{20}$ waveguide ($f$ = 13.9 GHz, $\mu_0 H_{ext}$ = 80 mT).

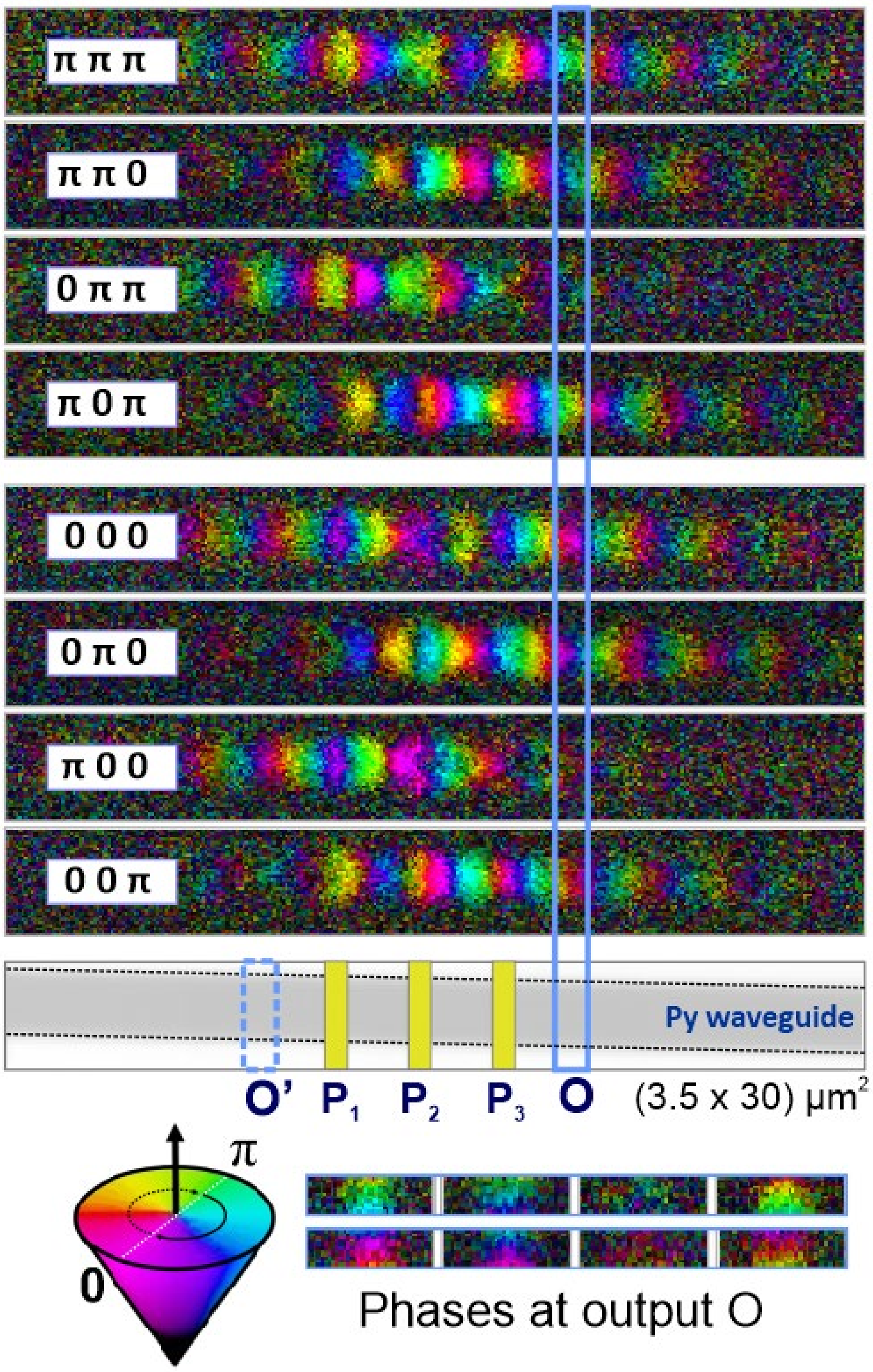

**Figure 2. Visualization of spin-wave majority gate operation.** Spatial phase distribution of the magnetization dynamics in a SWMG (2.0 μm wide permalloy waveguide, $f$ = 8.6 GHz, $\mu_0 H_{ext}$ = 80 mT) imaged by time-resolved scanning transmission x-ray microscopy for different sets of input phases. The spin-wave phase at the output position O allows for the reconstruction of the truth table of the majority function.

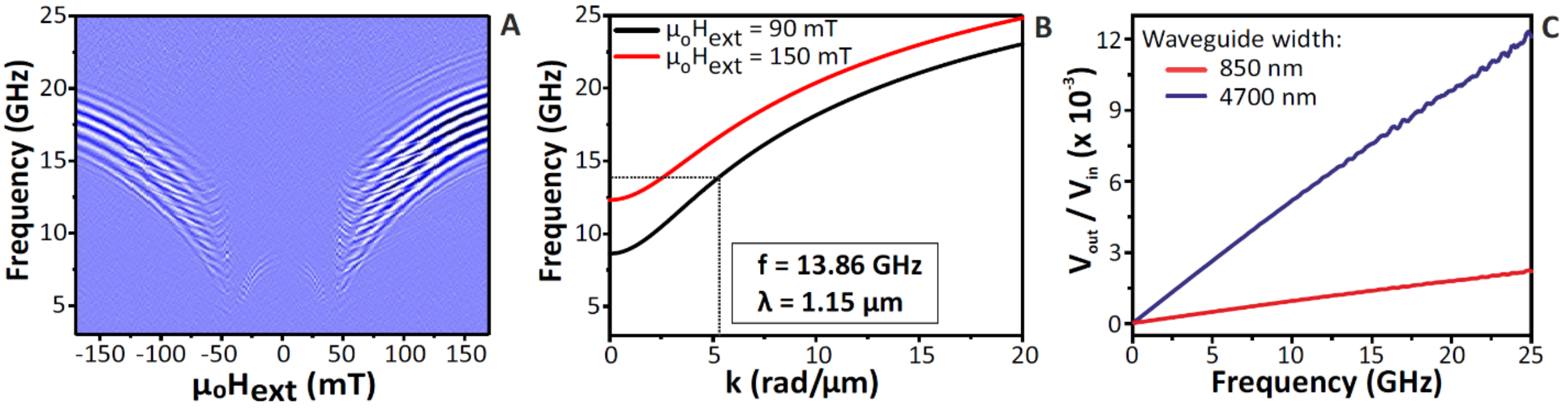

**Figure 3. Spin-wave dispersion and electrical operation of a nanoscale spin-wave majority gate. A.** Bias field derivative of the imaginary part of the $S_{21}$ microwave transmission parameter for spin-wave propagation in an 850 nm wide $Co_{40}Fe_{40}B_{20}$ waveguide between input $P_3$ and output O (distance of 6.9 μm) *vs.* frequency and transverse magnetic bias field. The oscillations stem from the phase accumulation during spin-wave propagation. **B.** Dispersion relation of spin waves in the Damon-Eshbach geometry in an 850 nm wide $Co_{40}Fe_{40}B_{20}$ waveguide for transverse magnetic bias fields as indicated. **C.** Voltage signal transmission ratio ($V_{in}/V_{out}$) determined from measured antenna impedances and micromagnetic simulations *vs.* spin-wave frequency.

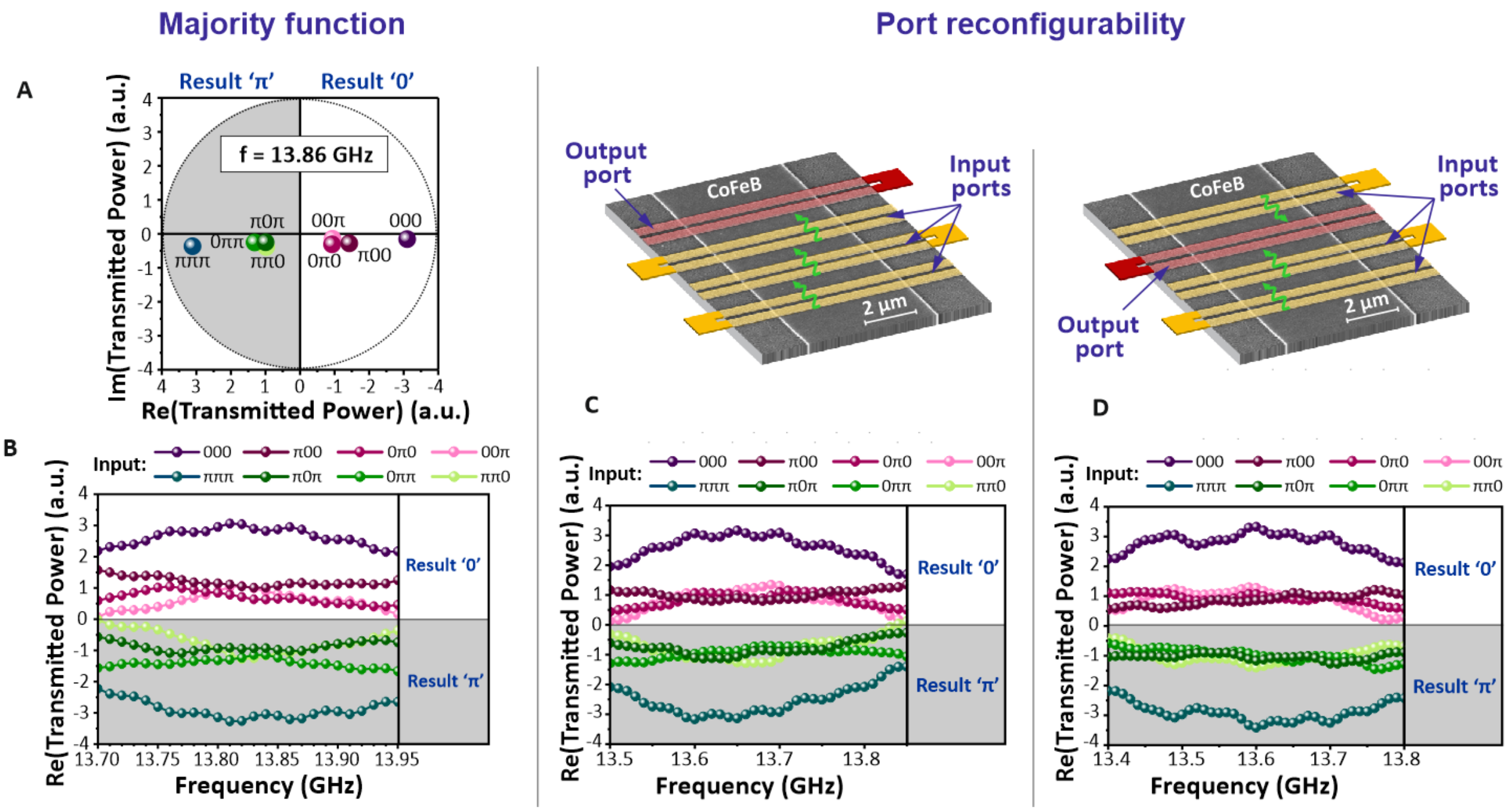

**Figure 4. All-electrical operation and reconfigurability of the spin-wave majority gate. A.** Polar plot of the transmitted power in a SWMG for different input phases under resonant conditions with $N = 2$ ($\mu_0 H_{ext}$ = 90 mT). Strong and weak majority signals can be clearly distinguished. **B.** Frequency-dependence of the real part of the transmitted power, indicating a 250 MHz wide frequency band for SWMG operation. **C.** Reconfigurability of the SWMG, configuration 1: schematic (top) and frequency-dependence of the real part of the transmitted power for different input phases in a SWMG using a 4.7-µm-wide $Co_{40}Fe_{40}B_{20}$ waveguide ($\mu_0 H_{ext}$ = 42 mT) and an outer antenna as output port. **D.** Configuration 2: experiment under identical conditions with an inner antenna as output port. In both cases, the majority gate operates over a frequency window larger than 300 MHz.

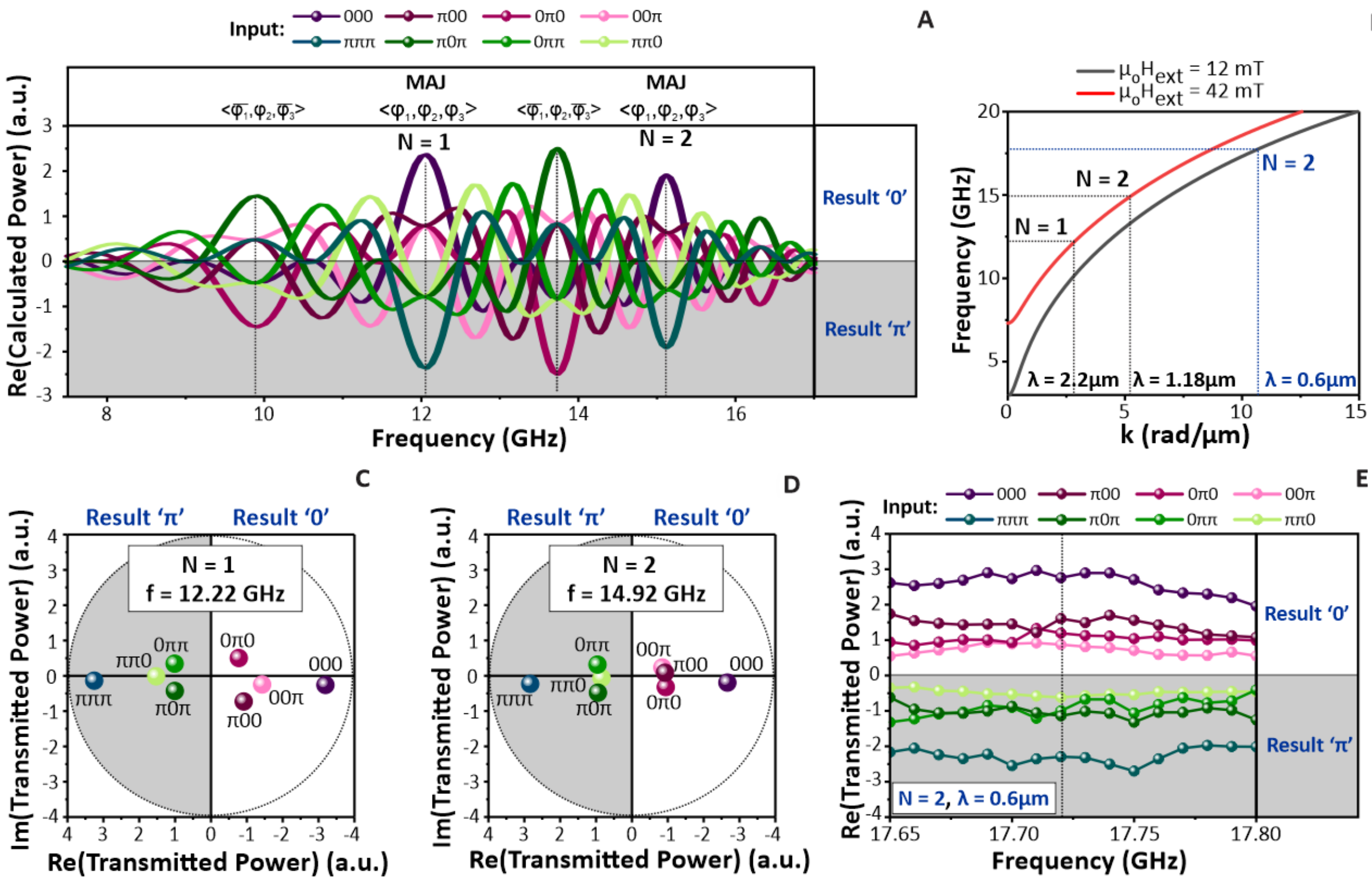

**Figure 5. Frequency-division multiplexing and operation at sub-µm wavelength. A.** Calculated real part of the output signal of a SWMG with a 4.7 µm wide $Co_{40}Fe_{40}B_{20}$ waveguide, a magnetic bias field of $\mu_0H_{ext}$ = 42 mT, and an interport spacing of 2.3 µm. **B.** Corresponding dispersion relations of spin waves in the Damon-Eshbach geometry for magnetic bias fields as indicated. **C** and **D** Complex polar plots of the transmitted power in the SWMG at $\mu_0H_{ext}$ = 42 mT and frequencies of 12.22 GHz ($N$ = 1) and 14.92 GHz ($N$ = 2), respectively. At both frequencies, the full truth table of the majority function is obtained in the same device. **E.** Frequency-dependence of the real part of the transmitted power as a function of the input phases in a SWMG with a 4.7 µm wide $Co_{40}Fe_{40}B_{20}$ waveguide, a magnetic bias field of $\mu_0H_{ext}$ = 12 mT, and an interport spacing of 1.2 µm. The majority function is obtained for a wavelength as low as 600 nm under resonant conditions with $N$ = 2.

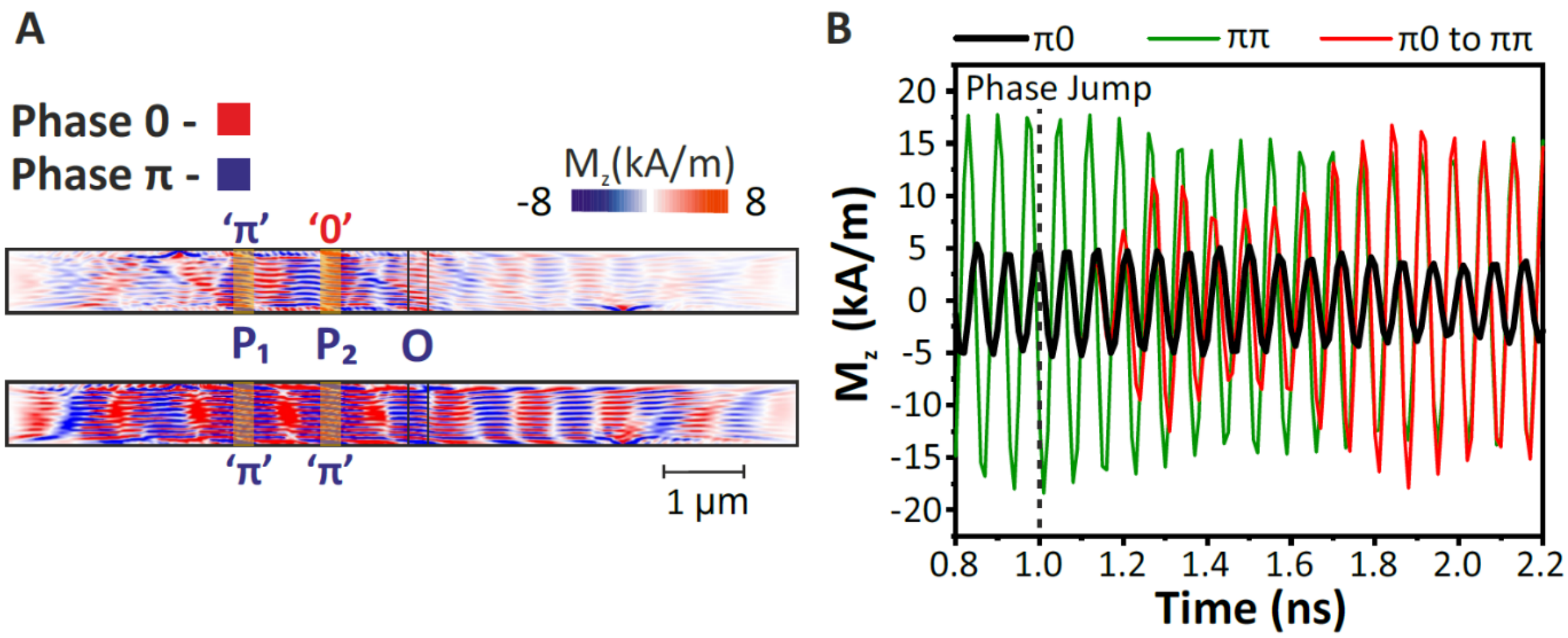

**Figure 6. Demonstration of sub-ns logic operation speed. A.** Snapshot images of the steady-state magnetization dynamics in an 850 nm wide $Co_{40}Fe_{40}B_{20}$ waveguide for (π0) or (ππ) input states ($f$ = 13.9 GHz, $\mu_0 H_{ext}$ = 80 mT). **B.** Magnetization oscillations at the output position O when the spin-wave signals at the two input ports $P_1$ and $P_2$ are either out-of-phase (π0) or in phase (ππ). In addition, the magnetization oscillation is shown when the spin-wave signal at port $P_2$ undergoes a phase jump of π (π0 to ππ) at 1.0 ns. The system adjusts to the new interference state within 700 ps.

**Table 1. Comparison of $Co_{40}Fe_{40}B_{20}$ and YIG as waveguide materials for the implementation of spin-wave majority gates.** The film thickness and the waveguide width are fixed to 850 and 30 nm, respectively, and an external field of $\mu_0 H_{ext}$ = 90 mT is applied transversally to the waveguide. In both cases, an operating wavelength of 1.15 µm is considered. The temperature dependence of the saturation magnetization is extracted from (35,36). The data show that $Co_{40}Fe_{40}B_{20}$ leads to much faster spin-wave propagation as well as much larger operational windows (frequency, temperature) than YIG.

| | $Co_{40}Fe_{40}B_{20}$ | YIG |
|---|---|---|
| Microstructure | Nanocrystalline | Single crystal epitaxial |
| Saturation magnetization (kA/m) | 1340 | 140 |
| Group velocity (µm/ns) | 5 | 0.5 |
| Attenuation length (µm) | 10 | 45 |
| Operating frequency (GHz) | 13.8 | 3.2 |
| SWMG Bandwidth (MHz) | 250–300 | 15 |
| Curie Temperature (K) | 1100 | 560 |
| Δf @FMR (MHz): T = 20°C to 100°C | 200 | 260 |

## Supplementary Materials

### *Animations of micromagnetic simulations (Movies S1 and S2)*

Animations of the magnetization dynamics in a SWMG with an 850 nm wide $Co_{40}Fe_{40}B_{20}$ waveguide calculated by micromagnetic simulations for different combinations of input phases: (0; 0; 0); (π; 0; π). The input phase combinations are indicated in the animations. Note that the complementary combinations (π; π; π) and (0; π; 0) can be obtained by shifting the time by half of a period.

#### *Movie S1*

Animated micromagnetic simulation of the magnetization dynamics in a SWMG with an 850 nm wide $Co_{40}Fe_{40}B_{20}$ waveguide for input phases of ***(0; 0; 0)***. Operation frequency 13.9 GHz, magnetic bias field $\mu_0 H_{ext}$ = 80 mT.

#### *Movie S2*

Animated micromagnetic simulation of the magnetization dynamics in a SWMG with an 850 nm wide $Co_{40}Fe_{40}B_{20}$ waveguide for input phases of ***(π; 0; π)***. Operation frequency 13.9 GHz, magnetic bias field $\mu_0 H_{ext}$ = 80 mT.

### *Animations of time-resolved scanning transmission x-ray microscopy (Movies S3, S4, and S5)*

Animations of the experimental magnetization dynamics in a SWMG with a 2.0 μm wide permalloy waveguide, imaged by time-resolved scanning transmission x-ray microscopy for different combinations of input phases: (π; π; π); (0; 0; π); (0; π; 0). The input phase combinations are indicated in the animations.

#### *Movie S3*

Experimental magnetization dynamics measured by time-resolved scanning transmission x-ray microscopy in a SWMG with a 2.0 μm wide permalloy waveguide for input phases of ***(π; π; π)***. Operation frequency 8.6 GHz, magnetic bias field $\mu_0 H_{ext}$ = 80 mT.

#### *Movie S4*

Experimental magnetization dynamics measured by time-resolved scanning transmission x-ray microscopy in a SWMG with a 2.0 μm wide permalloy waveguide for input phases of ***(0; 0; π)***. Operation frequency 8.6 GHz, magnetic bias field $\mu_0 H_{ext}$ = 80 mT.

***Movie S5***

Experimental magnetization dynamics measured by time-resolved scanning transmission x-ray microscopy in a SWMG with a 2.0 μm wide permalloy waveguide for input phases of ***(0; π; 0)***. Operation frequency 8.6 GHz, magnetic bias field $\mu_0 H_{ext}$ = 80 mT.

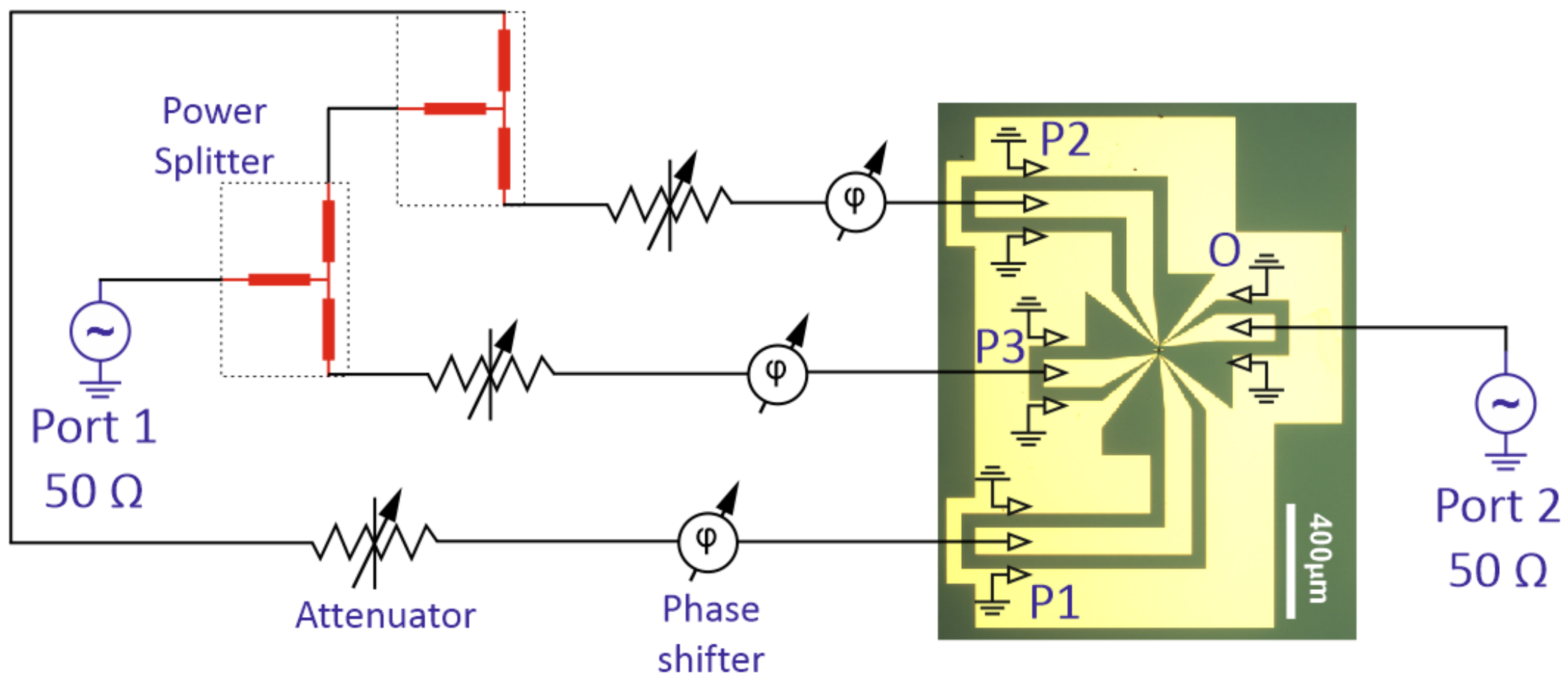

**Figure S1. Circuit schematic and optical micrograph of a SWMG device in an all-electrical experiment.** Power splitters, attenuators, and phase shifters used in the circuit to control the spin-wave properties are indicated, as well as the ground-signal-ground microwave probes connected to the coplanar waveguides used as contacts.

## Inductive microwave antenna design

Two or more inductive antennas can show strong parasitic coupling at microwave frequencies when they are at μm distances from each other and the power transmitted by such parasitic coupling can exceed the power transmitted by spin waves. Reducing their direct coupling in all-electrical experiment is fundamental to increase the ratio between the spin-wave signal and this "parasitic" transmission. To this aim, full-wave three-dimensional microwave simulations using the ANSYS HFSS (37) software were performed for the SWMG structures, without considering the magnetic properties of the ferromagnetic material. The centre-to-centre distance between adjacent antennas was 2.3 μm, as in the experimental devices. For comparison, a device implementing single wire antennas of 500 nm width was also simulated, similar to the devices used in the time-resolved scanning transmission x-ray microscope experiments. Fig. S2A shows the reflected power for the two antenna types. More importantly, the direct "parasitic" inductive coupling between two antennas was significantly reduced for the U-shaped *vs.* the single-wire antenna (Fig. S2B). As expected, the parasitic coupling was further progressively reduced when the distance between two U-shaped antennas was increased. This can be explained by the fact that the magnetic field generated by a single wire antenna decays with distance $r$ as $1/r$, while the field generated by a U-shaped antenna decays as $1/r^2$. The spatial distribution for the $x$ and $z$ components of the magnetic field generated by a 500 nm wide U-shaped antenna is shown in Fig. S3A; the corresponding spatial Fourier transforms in are shown in Fig. S3B for two different antenna widths, 500 and 250 nm. The spatial Fourier transform of the magnetic field can be expressed analytically by

$$\boldsymbol{F}(\boldsymbol{k}) = \mathbf{sinc}\left(\frac{\boldsymbol{kw}}{\mathbf{2}}\right) * \mathbf{sin}\left(\boldsymbol{k} * \frac{(\boldsymbol{w}+\boldsymbol{g})}{\mathbf{2}}\right), \tag{S1}$$

where $w$ is the width of the antenna and $g$ the gap of the U-shaped antenna. The spin-wave excitation efficiency is proportional to the magnitude of the spatial Fourier transform and is therefore zero for $k = 0$ (at ferromagnetic resonance) for a U-shaped antenna. Decreasing the width of the wires $w$ to 250 nm allows for more efficient excitation of spin waves with sub-μm wavelength, reaching wavelengths down to 600 nm, as shown in the paper.

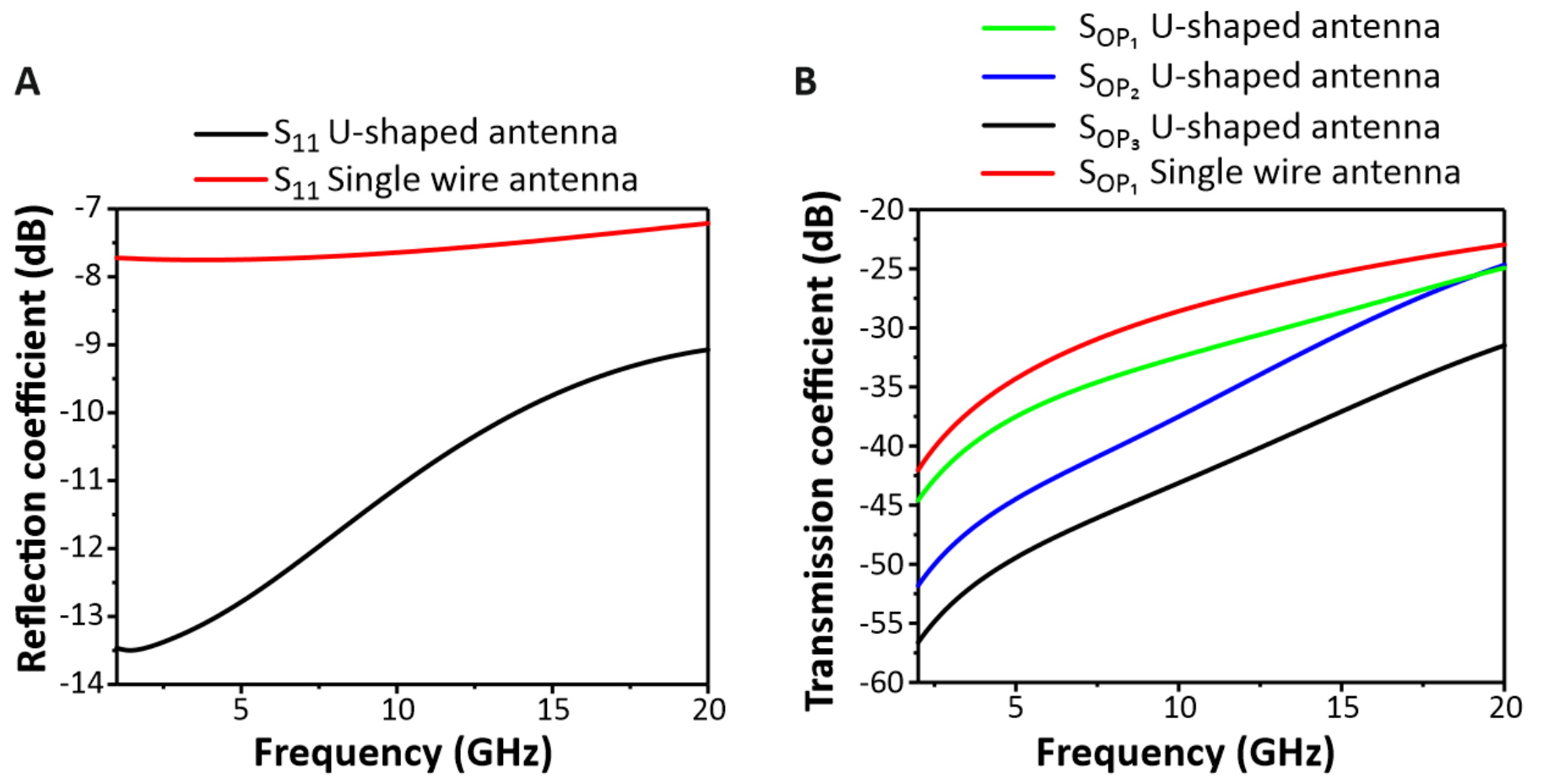

**Figure S2. Simulation of electrical properties of microwave antennas.** HFSS simulations of reflection (**A**) and transmission (**B**) S-parameters for U-shaped and single wire microwave antennas. The power transmission simulations in **B** for U-shaped antennas have been performed for distances of 2.3, 4.6, and 6.9 μm, corresponding to the distances between the input ports $P_1$, $P_2$, and $P_3$, and the output port O in the SWMG devices.

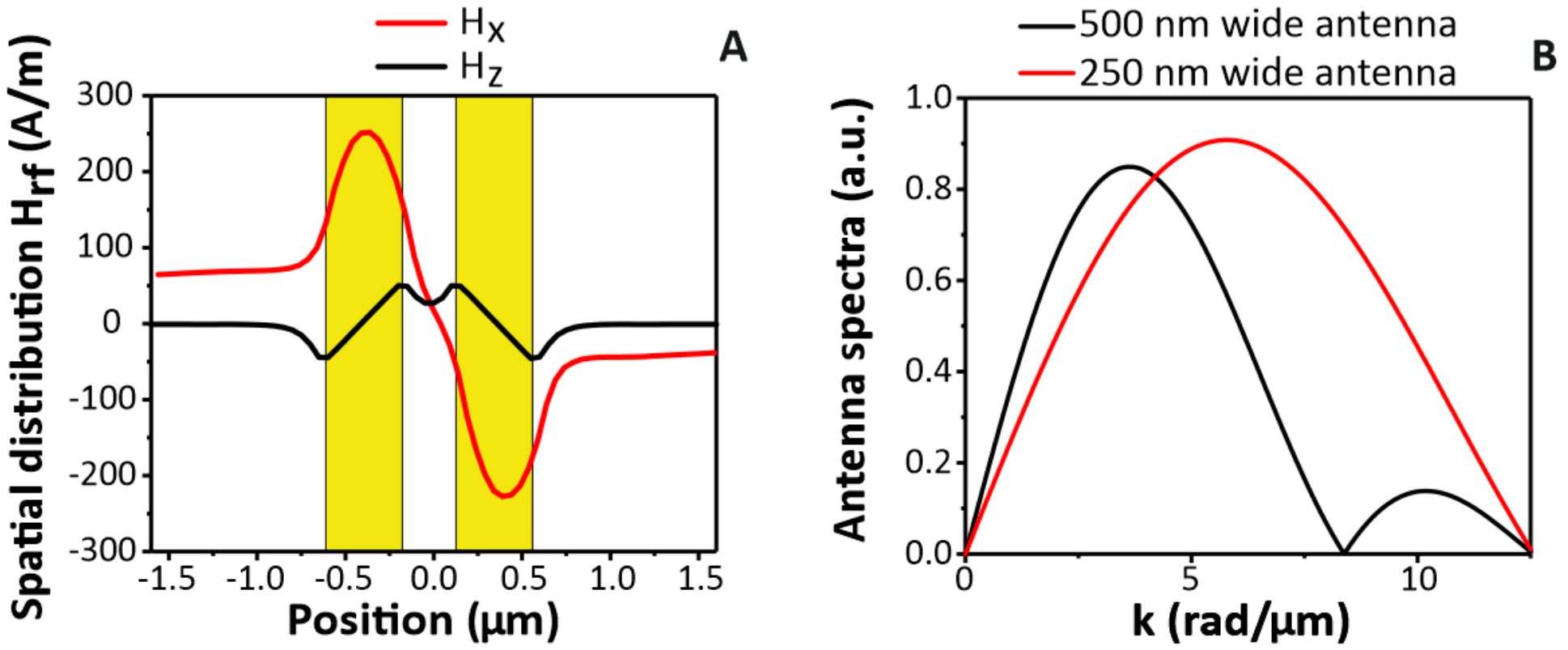

**Figure S3. Excitation field generated by microwave U-shaped antennas and its spectral content. A.** Magnetic field profile under a U-shaped antenna along the *x* and *z* direction. **B.** Spectra (spatial Fourier transforms) of the magnetic field representing the spin-wave excitation efficiency for the antennas with widths as specified.

**Analytical model of spin-wave transmission in all-electrical two-port experiments**

Calibrated two-port experiments of spin-wave transmission have been performed for different propagation distances as described in the paper. The resulting measurements are in good agreement with an analytical model considering the dispersion relation, the efficiency of spin-wave excitation by the antenna, and the phase accumulation of the spin wave during propagation.

The dispersion relation of spin waves confined in a transversally magnetized waveguide was calculated by the exchange-dipole model of Kalinikos and Slavin (24,38). In this analytical model, the frequency of a spin wave in a waveguide with width $w$ and thickness $d$ as a function of its wavevector $k$ is given by

$$\omega(\mathrm{k}) = \sqrt{(\omega_{\mathrm{H}} + \omega_{\mathrm{M}}\lambda_{\mathrm{ex}}[k^2 + (n\pi/w)^2])(\omega_{\mathrm{H}} + \omega_{\mathrm{M}}\lambda_{\mathrm{ex}}[k^2 + (n\pi/w)^2] + \omega_{\mathrm{M}}A)}, \qquad \text{(S2)}$$

with $\omega_{\mathrm{H}} = \gamma\mu_0 H_{\mathrm{eff}}$, $\omega_{\mathrm{M}} = \gamma\mu_0 M_s$, $\gamma$ the gyromagnetic ratio, $\mu_0 H_{\mathrm{eff}}$ the effective internal field, $M_s$ the saturation magnetization, $n$ (an integer) the mode number, and $\lambda_{\mathrm{ex}}$ the square of the magnetostatic exchange length. $A$ is given by

$$A = 1 - q\cos^2(\theta_{\mathrm{k}} - \theta_{\mathrm{M}}) + \frac{\omega_{\mathrm{M}} q(1-q)\sin^2(\theta_{\mathrm{k}} - \theta_{\mathrm{M}})}{\omega_{\mathrm{H}} + \omega_{\mathrm{M}}\lambda_{\mathrm{ex}}[k^2 + (n\pi/w)^2]}, \qquad \text{(S3)}$$

where $\theta_{\mathrm{k}} = \mathrm{atan}[n\pi/(kw)]$ is the angle between the spin-wave wavevector and the long axis of the waveguide, $\theta_{\mathrm{M}}$ is the angle between the magnetization direction and the long axis of the waveguide, and

$$q = 1 - \left[1 - \exp\left(-d\sqrt{k^2 + (n\pi/w)^2}\right)\right] / \left(d\sqrt{k^2 + (n\pi/w)^2}\right). \qquad \text{(S4)}$$

The excitation efficiency of a spin wave with wavevector $k$ is proportional to the magnitude of the Fourier component of the magnetic field, which is given by Eq. (S1) for U-shaped antennas. The phase accumulation during propagation is given by $\exp(ikr)$.

The experimental data and the results of the model are shown in Fig. S4 for waveguide widths of 4.7 µm and an external applied magnetic field of 42 mT. The parameters used for the analytical model were the same as used in the micromagnetic simulations reported in the paper: a saturation magnetization of 1.36 MA/m (determined experimentally), a Gilbert damping of 0.004 (determined experimentally by FMR), an exchange stiffness constant of 18.6 pJ/m, and a Landé factor of 2.07. Propagation distances corresponded to the geometry of the SWMGs used in the all-electrical experiments (2.3, 4.6, and 6.9 µm) and material parameters were those of $Co_{40}Fe_{40}B_{20}$ waveguide. The internal magnetic field was used as a fitting parameter since the combined effects of the demagnetising fields and the accuracy of the magnetic field calibration did not allow for the calculation of accurate dispersion relations based on nominal external magnetic bias fields. In the paper, nominal external field values are however mentioned for simplicity. For 4.7 µm wide $Co_{40}Fe_{40}B_{20}$ waveguides, the internal (fitted) magnetic field was about 36 mT, *i.e.* 6 mT less than external applied field of 42 mT, in agreement with the demagnetising field calculated by micromagnetic simulations. In this manner, good agreement between experiments and model could be obtained indicating the dispersion relation according to Eq. (S2) was accurate and efficient to calculate frequencies from wavelengths and to adjust resonant operation conditions of the SWMGs with 4.7 µm wide waveguides.

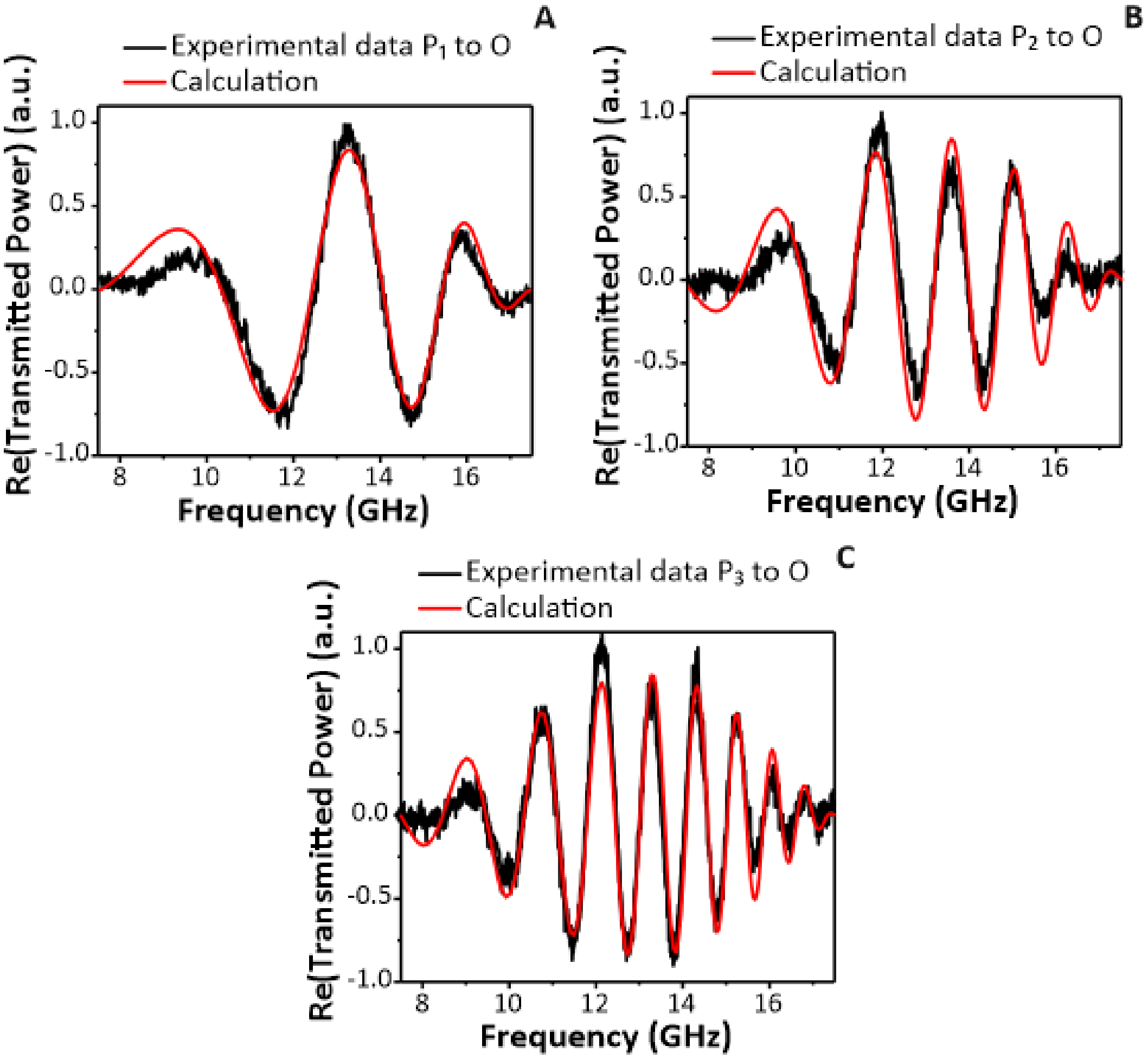

**Figure S4. Spin wave transmission in the 4.7 µm wide $Co_{40}Fe_{40}B_{20}$ waveguide.** Experimental spin wave transmission data and analytical calculations using the model described above for three different propagation distances of 2.3 µm (**A**), 4.6 µm (**B**), and 6.9 µm (**C**), respectively, in a majority gate device based on a 4.7 µm wide $Co_{40}Fe_{40}B_{20}$ waveguide.

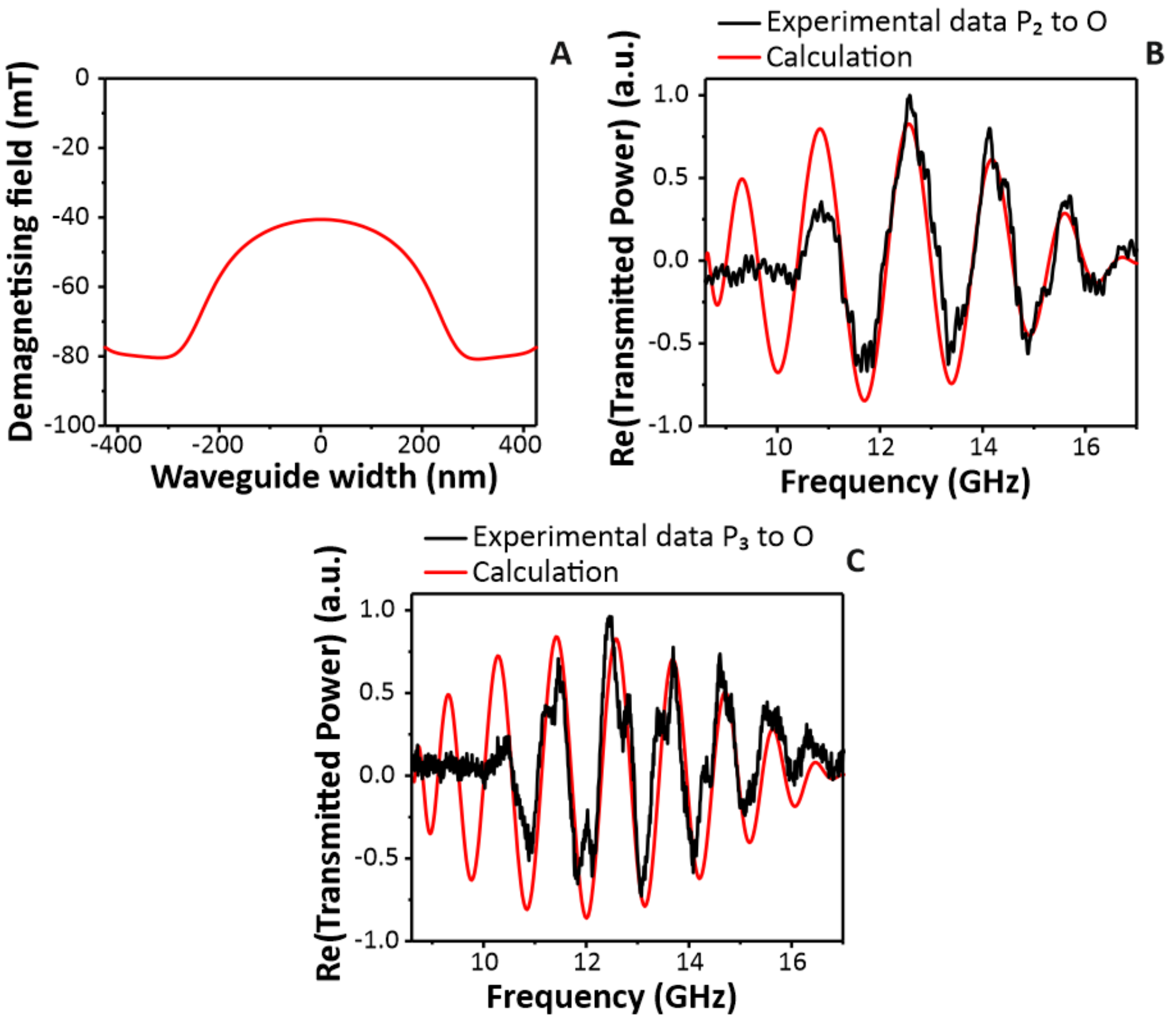

**Figure S5. Spin wave transmission in the 850 nm wide $Co_{40}Fe_{40}B_{20}$ waveguide. A.** Non-uniform demagnetising field across an 850 nm wide $Co_{40}Fe_{40}B_{20}$ waveguide calculated by micromagnetic simulations for an external magnetic bias field of 90 mT. **B** and **C.** Experimental spin-wave transmission data and analytical calculations using the above model for an 850 nm wide $Co_{40}Fe_{40}B_{20}$ waveguide and two different propagation distances, as indicated, showing good agreement.

For SWMGs with 850 nm wide waveguides, the internal field was found to be very inhomogeneous by micromagnetic simulations. Thus, spin waves are expected to propagate in an average effective magnetic bias field. The calculated demagnetising field was found to be 40 mT (*cf.* Fig. S5a) in the centre of an 850 nm wide $Co_{40}Fe_{40}B_{20}$, therefore the effective internal field is 50 mT for an external applied magnetic field of 90 mT. Towards the edges, the effective internal field is much lower. In addition, an effective width of the waveguide needed to be considered due to the inhomogeneity of the demagnetising field and the magnetization, which is nearly pointing along the waveguide at its edges under the chosen conditions. Good agreement between experiment and model was obtained for an effective width of 650 nm over the entire spin-wave band, as shown in Figs. S5B and S5C for propagation distances of 4.6 and 6.9 µm, indicating that the dispersion relation was accurate. An additional cut-off is also observed at large wavelengths (low frequencies), which can be attributed to the detection mechanism.

**Signal levels**

The signal level (voltage drop across the antenna) at an input port can be derived from applied input power (P = -6 dBm $\sim$ 0.25 mW) and the measured device impedance ($Z_{11}$), as

$$V_{in} = Z_{11}/(Z_{11}+R_s)\sqrt{2P(Rs+Z11)} \sim 120 \text{ mV}, \qquad \text{(S5)}$$

where $R_s$ is the series resistance of the VNA. Since the same antenna was used in all devices, the input voltage was the same in all experiments. The current flowing in the antenna is evaluated to be about 1.8 mA.

To estimate the voltage generated at an output antenna ($V_{out}$) due to a propagating spin wave, we consider a U-shaped antenna as a coil with a single turn and an area defined by the design and the waveguide width. The field generated by the spin waves was obtained from micromagnetic simulations using the experimental input current at a position of 60 nm above the waveguide surface, which corresponds to the experimental antenna position. The ratio between the calculated output voltage and the input voltage is shown as a function of frequency in Fig. 3C.

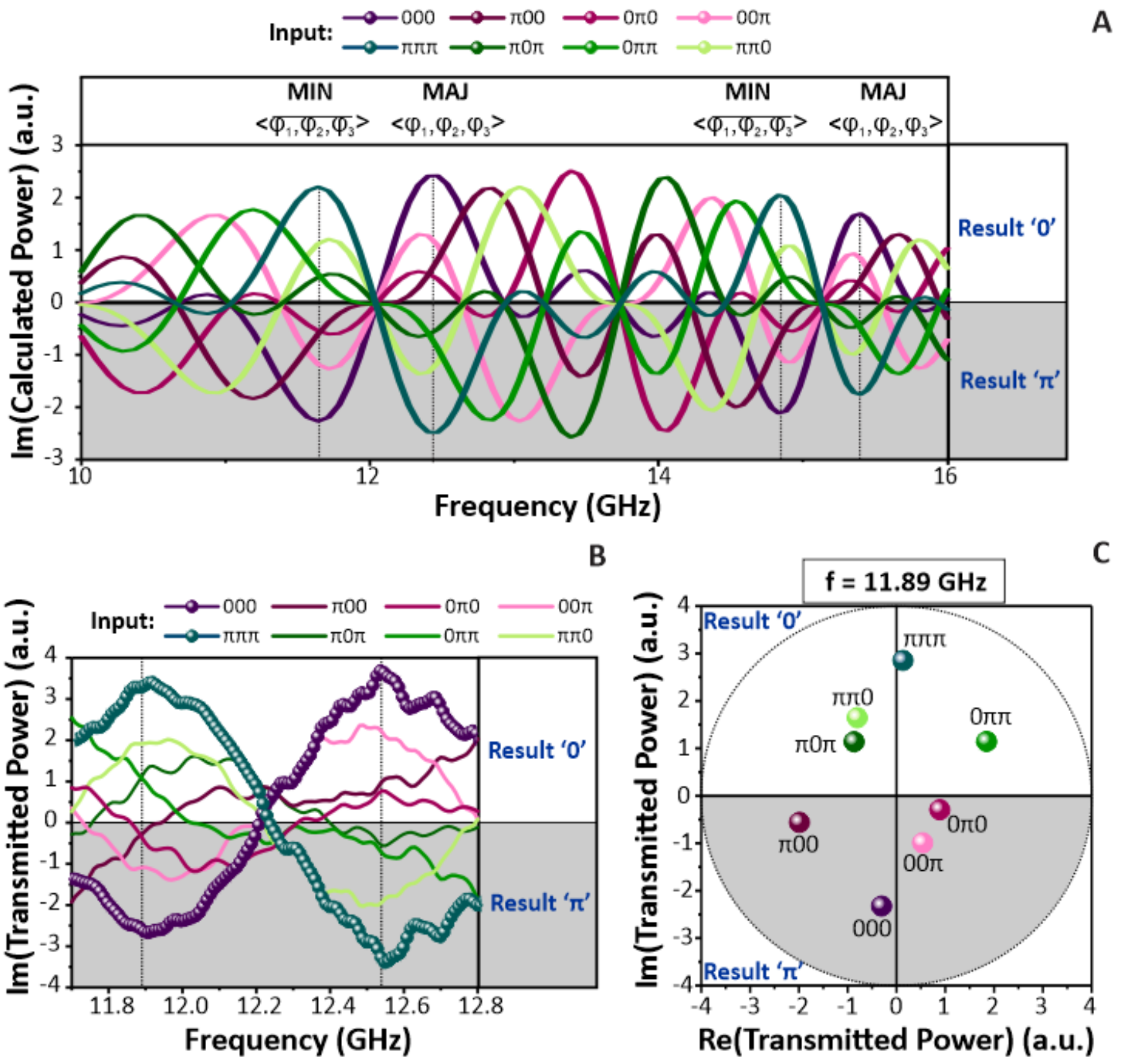

**Figure S6. Variants of logic functions: minority gate. A.** Calculated imaginary part of the output signal of a SWMG with a 4.7-μm-wide $Co_{40}Fe_{40}B_{20}$ waveguide, a magnetic bias field of $\mu_0 H_{ext}$ = 42 mT, and an interport spacing of 2.3 μm. Both MAJ and MIN (= ¬MAJ) functions are obtained in the same device. **B.** Experimental frequency dependence of the imaginary part of the transmitted power in the SWMG as a function of the input phases. MAJ and MIN logic functions are demonstrated at 12.53 GHz and 11.89 GHz, respectively. **C.** Complex polar plot of the transmitted power corresponding to minority gate operation at 11.89 GHz.

### Additional logic functions

Parallel majority and minority gate operation in a 16 $F^2$ minimum area SWMG is possible by taking advantage of the wave-based computing paradigm when the imaginary part of the signal—*i.e.* the signal that is shifted by $\pi/2$ with respect to the input reference phase of 0—is analyzed. For resonant majority gate conditions, the imaginary part of the transmitted signal is always zero, as shown in Fig. S6A at frequencies of 12.2 GHz and 15.1 GHz for $N = 1$ and $N = 2$, respectively. At nearby frequencies, conditions exist, where the wavelength is such that the phase accumulation during the propagation between adjacent transducers (distance of $2F$) is $2\pi \times N \pm \pi/4$. As shown in Fig. S6A, the imaginary part of the output signal then leads to the minority (majority) function of the inputs at the corresponding frequency below (above) the resonant one. This is experimentally confirmed in a SWMG with a 4.7 µm wide $Co_{40}Fe_{40}B_{20}$ waveguide, a magnetic bias field of $\mu_0 H_{ext} = 42$ mT, and an interport spacing of 2.3 µm for $N = 1$, operating the majority gate at 11.89 GHz ($\lambda = 2F/(1 - 1/8)$) and detecting the imaginary part of S6B and S6C. By contrast, the operation at 12.53 GHz ($\lambda = 2F/(1 + 1/8)$) leads to the truth table of the majority function (Fig. S6B). Again, higher harmonics can provide the same logic function and enable frequency-division multiplexing (Fig. S6A). These results show how the wave nature of the information carrier can be exploited to obtained parallel computation of identical or different logic functions in a single device.